\documentclass[12pt,oneside]{article}
\usepackage{amssymb}
\usepackage{amsmath}
\usepackage{amsfonts}
\usepackage[pagewise,mathlines]{lineno}
\usepackage{verbatim}
\usepackage{psfrag,graphicx}\graphicspath{{./figures/}}
\usepackage{amsmath,natbib}
\usepackage{setspace}
\usepackage{subfigure}
\usepackage{geometry,setspace}
\usepackage[page]{appendix}
\usepackage{color}
\newtheorem{theorem}{Theorem}

\newenvironment{proof}[1][Proof]{\noindent\textbf{#1.} }{\ \rule{0.5em}{0.5em}}
\newcommand{\bfI}{\mbox{\boldmath $I$}}

\newcommand{\ttheta}{\tilde \theta}
\newcommand{\apf}{auxiliary particle filter}

\newcommand{\MH}{Metropolis-Hastings}

\newcommand{\Uhat}{\widehat{U}}

\newcommand{\phat}{\widehat{p}}

\newcommand{\aimh}{adaptive independent Metropolis Hastings}

\newcommand{\mn}{mixture of normals}
\newcommand{\arwm}{adaptive random walk Metropolis}
\newcommand{\xtilde}{\widetilde {x}}
\newcommand{\pitilde}{\widetilde {\pi}}
\newcommand{\thetat}{\widetilde \theta}

\newcommand{\pf}{particle filter}
\newcommand{\ra}{\rightarrow}
\newcommand{\veps}{\varepsilon}

\newcommand{\up}[1]{\raisebox{0.25cm}[0pt]{#1}}
\title{\textbf{Particle filtering within adaptive \MH{} sampling}}

\author{Ralph S. Silva\\
        {\small School of Economics}\\
        {\small University of New South Wales}\\
        {\small r.silva@unsw.edu.au}
\and Paolo Giordani\\
        {\small Research Department}\\
        {\small Sveriges Riksbank}\\
        {\small paolo.giordani@riksbank.se}\\[0.5cm]
\and Robert Kohn\footnote{Corresponding author.}\\
        {\small School of Economics}\\
        {\small University of New South Wales}\\
        {\small r.kohn@unsw.edu.au}
\and Michael K. Pitt\\
        {\small Economics Department}\\
        {\small University of Warwick}\\
        {\small m.pitt@warwick.ac.uk}
}

\date{\vspace{0.5cm}{\small\today}}
\begin{document}
\maketitle
\thispagestyle{empty}
%--------------------------------------------------------------------------------------------------
% Abstract
%--------------------------------------------------------------------------------------------------
\begin{abstract}
 We show that it is feasible to carry out exact Bayesian inference for non-Gaussian state space models using an adaptive
\MH{} sampling scheme with the likelihood approximated by the \pf{}. Furthermore,  an \aimh{} sampler based on a \mn{}
proposal is computationally much more efficient than an \arwm{} proposal because the cost of constructing a good adaptive
proposal is negligible compared to the cost of approximating the likelihood. Independent \MH{} proposals are also
attractive because they are easy to run in parallel on multiple processors. We also show that when the particle filter is used,
the marginal likelihood of any model is obtained in an efficient and unbiased manner, making model comparison straightforward.

\end{abstract}

\newpage
\setstretch{2.0}
%--------------------------------------------------------------------------------------------------
%--------------------------------------------------------------------------------------------------
% Introduction
%--------------------------------------------------------------------------------------------------
%--------------------------------------------------------------------------------------------------
\section{Introduction}\label{section:introduction}
We show that it is feasible to carry out exact Bayesian inference on the parameters of a general state space model
by using the particle filter to approximate the likelihood and adaptive \MH{} sampling to generate unknown parameters.
The state space model can be quite general, but we assume that the observation equation can be evaluated analytically
and that it is possible to generate from the state transition equation.
Our methods are justified by the work
of \citet{andrieu:doucet:holenstein:2010} who show that the approximate likelihood is  the density of the
observations conditional on the parameters and a set of auxiliary uniform variables, with the states integrated out.

We consider a three component version of the \arwm{} proposal of
\cite{roberts:rosenthal:2008} and the \aimh{} proposal of  \citet{giordani:kohn:2008} which is based on a \mn{}
approximation to the posterior density. We show that the \aimh{} proposal can be much more efficient than the \arwm{} proposal
in terms of the computing time required to achieve a given level of accuracy for three reasons.
The first reason is that it is important to construct efficient adaptive proposals because the approximate likelihood is stochastic and not a smooth function of the parameters \citep[see][]{pitt:2002}. This means that small changes in the
parameters can result in large changes in the approximate likelihood so that a sampling scheme such as a
random walk that changes the parameters by small amounts to try and obtain adequate acceptance may not work well in this context.
Second, it is  worthwhile constructing efficient adaptive proposals because
the cost of the adaptation steps is negligible compared to the cost of approximating the marginal likelihood using the particle
filter. The high cost of approximating the likelihood occurs
as it is necessary to use a large number of particles to obtain an adequate approximation and it is necessary to run the \pf{}
thousands of times for simulation based
inference. Third, it is much easier to run an \aimh{} scheme in parallel on multiple processors
than an \arwm{} scheme and such parallel processing can reduce computational time significantly
for a given level of accuracy; in many of our examples  the reduction is by a factor of five to thirty
when running in parallel on eight processors.

Our article also shows that when particle filtering is used,
 the marginal likelihood of any  model can be obtained using bridge sampling or importance sampling in an efficient and unbiased manner making   model comparison straightforward. The methodology is illustrated
empirically using challenging models and data.

Adaptive sampling methods are simulation methods for carrying out Bayesian inference that use previous iterates of the simulation
 to form proposal distributions, that is, the adaptive samplers learn about the posterior distribution from previous iterates.
 See for example \cite{haario:saksman:tamminen:2001}, \cite{atchade:rosenthal:2005} and  \cite{roberts:rosenthal:2008}
 who consider \arwm{} proposals and \cite{giordani:kohn:2008} who base their proposal on a \mn{}. Adaptive sampling is
 particularly attractive when the \pf{} is used to approximate the posterior density because it is difficult to form proposal
 densities by constructing approximations that require derivatives of the log likelihood.

Particle filtering (also known as sequential Monte Carlo) was first proposed by \cite{gordon:salmond:smith:1993} for online filtering and prediction of nonlinear or non-Gaussian state space models.
The auxiliary particle filter method was introduced by  \cite{pitt:shephard:1999} to improve the performance of the standard particle filter when the observation equation is informative relative to the state equations, that is when the signal to noise ratio is moderate to high.  There is an extensive literature on online filtering  using the particle filter, see for example \cite{kitagawa:1996}, \cite{liu:chen:1998}, \cite{doucet:godsill:andrieu:2000}, \cite{doucet:freitas:gordon:2001}, \cite{andrieu:doucet:2002}, \cite{fearnhead:clifford:2003} and \cite{delmoral:doucet:jasra:2006}. Our article considers only the standard particle filter of \cite{gordon:salmond:smith:1993}  and the generic auxiliary particle filter of \cite{pitt:shephard:1999}.

The literature on using the particle filter to learn about model parameters is more limited.
\cite{pitt:2002} proposes the smooth \pf{} to estimate the parameters of a state space using maximum likelihood. \cite{storvik:2002} and \cite{polson:stroud:muller:2008} consider online parameter learning when sufficient statistics are available.  \citet{andrieu:doucet:holenstein:2010} provide a framework for off line parameter learning using the \pf{}.
 \citet{flury:shephard:2008} give an insightful discussion of the results of \citet{andrieu:doucet:holenstein:2010}
 and use single parameter random walk proposals to carry out off-line Bayesian inference.

%--------------------------------------------------------------------------------------------------
%--------------------------------------------------------------------------------------------------
% Particle filter
%--------------------------------------------------------------------------------------------------
%--------------------------------------------------------------------------------------------------
\section{State space models}\label{section:PF}
Consider a state space model with observation equation $p(y_t|x_t;\theta)$ and state transition equation
$p(x_t|x_{t-1};\theta)$, where $y_t$ and $x_t$ are the observation and the state at time $t$ and $\theta$
is a vector of unknown parameters. The distribution of the initial state is
$p(x_0|\theta)$.  {See \citet{cappe:moulines:ryden:2005} for a modern treatment of general state space models}.
The filtering equations for  the state space model (for $t \geq 1$) are
\citep[pp.~506-507]{west:harrison:1997}
\begin{subequations}
\begin{align}
p(x_t|y_{1:t-1};\theta) & =\int p(x_t|x_{t-1};\theta)p(x_{t-1}|y_{1:t-1};\theta)dx_{t-1},  \label{eq: update1}\\
p(x_t|y_{1:t};\theta)&= \dfrac{p(y_t|x_{t};\theta)p(x_{t}|y_{1:t-1};\theta)}{p(y_t|y_{1:t-1};\theta)}, \label{eq: update2}\\
p(y_t|y_{1:t-1};\theta) &= \int p(y_t|x_{t};\theta)p(x_{t}|y_{1:t-1};\theta)dx_t \label{eq: one term likelihood}.
\end{align}
\end{subequations}
where $y_{1:t} = \{y_1, \dots, y_t\}$.  Equations~\eqref{eq: update1}--\eqref{eq: one term likelihood}  allow
(in principle) for filtering for a given $\theta$ and for evaluating the likelihood of the observations $y = y_{1:T}$,
\begin{align}\label{eq: likelihood}
p(y|\theta) & = \prod_{t=0}^{T-1} p(y_{t+1}|y_{1:t};\theta)\ ,
\end{align}
where $y_{1:0}$ is a null observation.
If the likelihood $p(y|\theta)$ can be computed, maximum likelihood and MCMC methods can be used to carry out inference
on the parameters $\theta$, with the states integrated out.
When both the observation and state transition equations are linear and Gaussian the likelihood can be
evaluated analytically using the Kalman filter
\citep[pp.~141-143]{cappe:moulines:ryden:2005}. More general state space models can also be estimated
by MCMC methods if auxiliary variables are introduced, e.g. \citet{kim:shephard:chib:1998} and \citet{fruhwirhschnatter:wagner:2006} and/or the states are sampled in blocks
as in \citet{shephard:pitt:1997}. See Section 6.3 of \citet{cappe:moulines:ryden:2005} for a review of Markov chain Monte Carlo methods applied to general state space models. In general, however, the integrals in equations
\eqref{eq: update1}--\eqref{eq: one term likelihood} are computationally intractable and the standard particle filter is proposed by \cite{gordon:salmond:smith:1993} as a method for approximating them with
the approximation becoming exact as the number of particles tends to infinity. Appendix~\ref{A: regular PF}
describes the standard particle filter and its  use in
approximating the expressions in \eqref{eq: update1}--\eqref{eq: one term likelihood}. We refer to \cite{pitt:shephard:1999}
for a description of the auxiliary \pf{} and \cite{pitt:2002} for the efficient computation of the likelihood based on the
auxiliary \pf{}.

%--------------------------------------------------------------------------------------------------
%--------------------------------------------------------------------------------------------------
% Regular particle filter
%--------------------------------------------------------------------------------------------------
%--------------------------------------------------------------------------------------------------

%--------------------------------------------------------------------------------------------------

%--------------------------------------------------------------------------------------------------
\section{Adaptive sampling}\label{section:adaptive:sampling}
Suppose that $\pi(\theta)$ is the target density from which we wish to generate a sample, but that it is computationally difficult to do so directly. One way of generating the sample is to use the Metropolis-Hastings method, which is now described. Suppose that given some initial $\theta_0$ the $j-1$ iterates $\theta_1, \dots, \theta_{j-1}$ have been generated. We then generate $\theta_{j}$ from the proposal density $q_{j}(\theta;\thetat) $ which may also depend on some other value of $\theta$ which we call $\thetat$. Let $\theta_{j}^p$ be the proposed value of $\theta_{j}$ generated from $q_{j}(\theta;\theta_{j-1}) $. Then we take $\theta_{j} = \theta_{j}^p$ with probability
\begin{align} \label{e:adaptive accep prob}
\alpha(\theta_{j-1};\theta_{j}^p) = \min \biggl \{1,\frac{\pi(\theta_{j}^p)}{\pi(\theta_{j-1})}
\dfrac{q_{j}(\theta_{j-1};\theta_j^p)}{q_{j}(\theta_j^p;\theta_{j-1})}\biggr \} \ ,
\end{align}
and take $\theta_j = \theta_{j-1}$ otherwise. If $q_j(\theta;\thetat)$ does not depend on $j$, then under appropriate regularity conditions we can show that the sequence of iterates $\theta_j$ converges to draws from the target density $\pi(\theta)$. See \cite{tierney:1994} for details.

In adaptive sampling the parameters of $q_j(\theta;\thetat)$ are estimated from the iterates $\theta_1, \ldots, \theta_{j-2}$. Under appropriate regularity conditions  the sequence of iterates $\theta_j, j \geq 1$, converges to draws from the target distribution $\pi(\theta)$. See \cite{roberts:rosenthal:2007}, \cite{roberts:rosenthal:2008} and \cite{giordani:kohn:2008}.

In our applications the target distribution is $p(\theta|y)$  is not available in a known closed form, but
the standard and auxiliary particle filters provide unbiased estimates of the likelihood function \citep{delmoral:2004}.
\citet{andrieu:doucet:holenstein:2010} show that we can view the particle filter approximation to the likelihood $\phat (y|\theta)$ as the density of $y$ conditional on $\theta $ and a set of auxiliary uniform variables $u$ such that
$\phat (y|\theta) = f(y|\theta, u ) $ and
\begin{align} \label{eq: unbiassed lik}
\int f(y|\theta,u)f(u|\theta)du  & = p(y|\theta).
\end{align}
It follows that $f(\theta|y)  = p(\theta|y)$ so that a method that simulates from $f(\theta,u|y)$ yields iterates from the correct posterior $p(\theta|y)$. In particular an adaptive sampling method using the particle filter to estimate the likelihood
can be considered as an auxiliary variable method to sample from the augmented target $p(\theta,u|y)$
such that the joint proposal distribution for $\theta $ and $u$ is $q(\theta,u; \thetat) = q(\theta;\thetat)p(u|\theta)$
with $u$ a vector of uniform variables. The acceptance probability \eqref{e:adaptive accep prob} for an adaptive proposal
$q_j(\theta,u; \thetat)$ becomes
\begin{align} \label{e:adaptive accep prob pf}
\alpha(\theta_{j-1},u_{j-1};\theta_{j}^p,u^{p}) = \min \biggl \{1,\frac{p(y|\theta_{j}^p,u_j^p)p(\theta^p)}
{p(y|\theta_{j-1},u_{j-1})p(\theta_{j-1})}
\dfrac{q_{j}(\theta_{j-1};\theta_j^p)}{q_{j}(\theta_j^p;\theta_{j-1})}\biggr \} \ .
\end{align}
If the adaptive proposal is independent, i.e. $q_j(\theta,u; \thetat) = q_j(\theta,u)$, then
\begin{align} \label{e:adaptive indep accep prob pf}
\alpha(\theta_{j-1},u_{j-1};\theta_{j}^p,u^{p}) = \min \biggl \{1,\frac{p(y|\theta_{j}^p,u_j^p)p(\theta^p)}
{p(y|\theta_{j-1},u_{j-1})p(\theta_{j-1})}
\dfrac{q_{j}(\theta_{j-1})}{q_{j}(\theta_j^p)}\biggr \} \ .
\end{align}

The two adaptive sampling schemes studied in the paper are discussed in appendix~\ref{app: adaptive sampling}.

The following convergence results hold for the \aimh{}
sampling scheme described in appendix~\ref{subsection:AIMH:MN} (and more fully in \cite{giordani:kohn:2008})
when it is combined with the standard particle filter. They  follow from Theorems 1 and 2 of \cite{giordani:kohn:2008}.
Let $\Theta$ be the parameter space of $\theta$.
\begin{theorem} \label{thm: convergence}
Suppose that (i)~$p(y_t|x_t;\theta) \leq \phi_t$ for $t=1,\dots, T,$ where $\phi_t$
is functionally independent of $\theta\in \Theta$ and
$x_t$ and (ii)~$p(\theta)/g_2(\theta) \leq C$  for any $\theta \in \Theta$ where $C$ is a constant and
the density $g_2(\theta)$ is the second component in the mixture proposal. Then,
\begin{enumerate}
\item
The iterates $\theta_j$ of the \aimh{} sampling scheme converge to a sample from $p(\theta|y)$ in the sense that
\begin{align} \label{e:theorem1}
\sup_{A \subset \Theta} \mid \Pr(\theta_j  \in A) - \int_A p(\theta \mid y )d\theta  \mid & \ra 0 \quad \text{as} \quad j \ra \infty.
\end{align}
for all measurable sets $A$ of $\Theta$.
\item
Suppose that $h(\theta)$ is a measurable function of $\theta$
that is square integrable with respect to the density $g_2$. Then, almost surely,
\begin{align} \label{e:theorem2}
\frac{1}{n} \sum_{j=1}^n h (\theta_j) \ra \int h(\theta)p(\theta|y) d\theta  \quad  \text{as   } \quad n \ra
\infty.
\end{align}
\end{enumerate}
\begin{proof}
\begin{align*}
\phat (y|\theta) & = \prod_{t=0}^{T-1} \phat (y_{t+1}|y_{1:t};\theta)
 \leq \prod_{t=1}^T \phi_t \qquad \text{ because} \\
\phat (y_t|y_{1:t-1};\theta) & = \frac1M \sum_{j=1}^M p(y_t|x_t^j; \theta )  \leq \phi_t
\end{align*}
by \eqref{eq:RPF:likelihood}. This shows that the approximate likelihood is bounded and the result now follows from
\cite{giordani:kohn:2008} when we make the second component heavy tailed compared to the prior, as outlined in that paper.
\end{proof}
\end{theorem}
The theorem applies to the stochastic volatility, negative binomial and Poisson state space models discussed in section~\ref{section:algorithm:comparison} as well as to binary and binomial state space models.

We can obtain a similar convergence result for the \apf{} if it is modified in a straightforward way to ensure that
the importance weights are bounded.  The proof is outlined in appendix~\ref{A: APF}
\begin{theorem}\label{thm: convergence apf}
Subject to the conditions of theorem~\ref{thm: convergence} and the construction of the importance weights
in appendix~\ref{A: APF}, the results of
theorem~\ref{thm: convergence} also hold for the \apf{}.
\end{theorem}

%--------------------------------------------------------------------------------------------------
%--------------------------------------------------------------------------------------------------

\subsection{Adaptive sampling and parallel computation}\label{ss: AS and parallel computation}
 Our work uses parallel processing for adaptive sampling in two ways.
 Suppose  $J$ processors are available. The first approach applies to any sampling scheme.
 The likelihood is estimated for a given $\theta$ on each of the processors using the particle filter with $M$ particles and these estimates are then averaged to get an estimate of the likelihood based on $JM$ particles. This approach is similar to, but faster, than using a single processor and makes it possible to estimate the likelihood using a large number of particles.

The second approach applies mainly to independent \MH{} sampling schemes and
consists of iterating on the following three steps. Let $\theta^c$ the current value of $\theta$ generated by the sampling scheme and $q_c(\theta)$ the current proposal density for $\theta$. (a)~For each of $J$ processors generate
 $K$ proposed values of $\theta$, which we write as $\theta^{(p)}_{j,k},k=1, \dots, K$,
and compute the corresponding logs of the
 ratios $ \phat
 (y|\theta^{(p)}_{j,k})p(\theta^{(p)}_{j,k}) /q(\theta^{(p)}_{j,k})$.
 (b)~After each $K$ block of proposed values is generated for each processor,
 carry out \MH{} selection of
 the $JK$ proposed $\{\theta^{(p)}_{j,k}\}$ parameters using a single processor to
 obtain $\{\theta_{j,k}\}$ draws from the chain.
 This is fast because drawing uniform variates is the  only computation that is necessary.
(c)~Use the previous iterates and the $\theta_{j,k}$ to update the proposal density $q_c(\theta)$ and $\theta_c$.

%--------------------------------------------------------------------------------------------------
%--------------------------------------------------------------------------------------------------
% Bridge sampling
%--------------------------------------------------------------------------------------------------
%--------------------------------------------------------------------------------------------------
\subsection{Estimating the marginal likelihood}\label{subsection:bridge:sampling}
Marginal likelihoods are often used to compare two or more models. For a given model, let $\theta$ be the vector of model parameters, $p(y|\theta)$ the likelihood of the observations $y$ and $p(\theta)$ the prior for $\theta$. The marginal likelihood is
\begin{align}\label{eq:marg:likelihood}
p(y) & =\int p(y|\theta)p(\theta)d\theta.
\end{align}
which in our case becomes
\begin{align}\label{eq:marg:likelihood pf}
p(y) & =\int p(y|\theta,u)p(\theta)p(u) d\theta\ du.
\end{align}
It is often difficult to evaluate or estimate $p(y)$ and appendix~\ref{A: BS and IS} briefly outlines how it can be estimated
using bridge and importance sampling, with the computation carried out within the adaptive sampling
so that a separate simulation run is unnecessary.

%--------------------------------------------------------------------------------------------------
%--------------------------------------------------------------------------------------------------
% Algorithm comparison
%--------------------------------------------------------------------------------------------------
%--------------------------------------------------------------------------------------------------
\section{Performance of the adaptive sampling schemes}\label{section:algorithm:comparison}
This section compares the performance of the two adaptive \MH{} sampling schemes discussed in
section~\ref{section:adaptive:sampling} using both the standard particle filter and the auxiliary particle filter.
The comparisons are carried out for several models using real data and illustrate the flexibility and wide applicability of the
approach that combines particle filtering with adaptive sampling.
The comparison is in terms of the acceptance rates of the Metropolis-Hastings methods, the
inefficiency factors (IF) of the parameters, and an overall measure of effectiveness which compares the times taken by each
combination of sampler and particle filter  to obtain the same level of accuracy. We define the acceptance rate as the percentage
of accepted values of each of the Metropolis-Hastings proposals. We define the inefficiency of the sampling scheme for a given
parameter as the variance of the parameter estimate divided by its variance if the sampling scheme generates independent iterates. We estimate the inefficiency factor as
%\begin{equation}
$\text{IF} = 1+2\sum_{j=1}^{L}\hat{\rho}_j,$
%\label{eq:ineffic}
%\end{equation}
where $\hat{\rho}_j$ is the estimated autocorrelation at lag $j$. As a rule of  thumb, the maximum number of lags $L$ that
we use is given by the lowest index $j$ such that $|\hat{\rho}_j|<2/\sqrt{K}$ where $K$  is
the sample size used to compute $\hat{\rho}_j$. The acceptance rate and the inefficiency factor do not take into account the time taken by a sampler. To obtain an overall measure of the effectiveness of a sampler, we define its equivalent computing time $ ECT = 10 \times IF \times t$, where $t$ is the time per iteration of the sampler. We interpret $ECT$ as the time taken by the sampler to attain the same accuracy as that attained by 10 independent draws of the same sampler. For two samplers $a$ and $b$,  $ECT_a/ECT_b$ is the ratio of times taken by them to achieve the same accuracy.

We note that the time per iteration for a given sampling algorithm depends  on how the algorithm is implemented, i.e.
the language used, whether operations are vectorized, etc. Thus the implementation of the sampling scheme affects its ECT, but  not the acceptance rates nor the inefficiencies. Implementation details are given in appendix~\ref{app: Implementation}.

%--------------------------------------------------------------------------------------------------
%--------------------------------------------------------------------------------------------------
% Stochastic volatility model
%--------------------------------------------------------------------------------------------------
%--------------------------------------------------------------------------------------------------
\subsection{Example~1: Stochastic volatility model} \label{ss: sv model}
The first example considers the univariate stochastic volatility (SV) model
\begin{align}\label{eq: sv general}
\begin{split}
y_{t}      & = K_t\exp(x_t/2)\varepsilon_t,\hspace{2cm}\varepsilon_t\sim\mathcal{N}(0,1) \\
x_{t} & =  \mu + \phi(x_{t-1}-\mu)+\sigma_\eta\eta_t,\hspace{0.6cm}\eta_t\sim\mathcal{N} (0,1)
%y_{t}      & = K_t\exp(\alpha_t/2)\varepsilon_t, &  \varepsilon_t\sim\mathcal{N}(0,1) \\
%\alpha_{t} & =  \mu + \phi(\alpha_{t-1}-\mu)+\sigma_\eta\eta_t,  & \eta_t\sim\mathcal{N} (0,1)
\end{split}
\end{align}
where $\text{corr}(\varepsilon_t,\eta_t)=\rho$, $\Pr(K_t=2.5)=\omega$ and $\Pr(K_t=1)=1-\omega$, with $\omega << 1$. This is a state space model with a non-Gaussian observation equation  and a Gaussian state transition equation for the latent volatility $x_t$ which follows a first order autoregressive model. The SV model allows for leverage because the errors in the observation and state transition equations can be correlated. The model also allows for outliers in the observation equation because the standard deviation of $y_t$ given $x_t$ can be 2.5 its usual size when $K_t = 2.5$. To complete the model specification,  we assume that all parameters are independent a priori with the following prior distributions: $\mu\sim\mathcal{N}(0,10)$, $\phi\sim\mathcal{TN}_{(0,1)}(0.9,0.1)$, $\sigma_\eta^2\sim\mathcal{IG}(0.01,0.01)$, $x_0\sim\mathcal{N}(0,10)$, and $\rho\sim\mathcal{TN}_{(-1,1)}(0,10^6)$ where $\mathcal{N}(a,b)$ means a normal distribution with mean $a$ and variance $b^2$, $\mathcal{TN}_{(c,d)}(a,b)$ means a truncated normal with location $a$ and scale $b$
 restricted to the interval $(c,d)$  and $\mathcal{IG}(a,b)$ is an inverse gamma distribution with shape parameter $a$, scale parameter $b$ and mode $b/(a+1)$. We set $\omega=0.03$ in the general model.

\cite{shephard:2005} reviews SV models and a model of the form \eqref{eq: sv general} is estimated by \citet{malik:pitt:2008}
by maximum likelihood using the smooth \pf{}.

%--------------------------------------------------------------------------------------------------
%--------------------------------------------------------------------------------------------------
% Stochastic Volatility model: S & P 500 data set
%--------------------------------------------------------------------------------------------------
%--------------------------------------------------------------------------------------------------
\subsubsection{S\&P 500 index}\label{sss: sp}
We apply the  SV model \eqref{eq: sv general} to the Standard and Poors (S\&P) 500 data from 02/Jan/1970 to 14/Dec/1973 obtained from Yahoo Finance web site\footnote{ http://au.finance.yahoo.com/q/hp?s=\^{}GSPC}. The data consists of $T=1~000$ observations.

Table~\ref{table:SP500:data:MC:study} presents the results of a Monte Carlo study using twelve replications with different random number seeds  for the SV model  with leverage using the first parallel computation method described in \ref{ss: AS and parallel computation}. Implementation details are given in
appendix~\ref{app: S&V implement details}. The table shows that the adaptive independent Metropolis-Hastings sampling scheme is at least seven times more efficient than the adaptive random walk Metropolis sampling scheme for both particle filters.

%\begin{sidewaystable}[!ht]
\begin{table}[!ht]
\centering
\caption{Medians and interquartile range (IR) of the acceptance rates and the inefficiencies (minimum, median and maximum) and ECT = IF $\times$ time for ten iterations over twelve replications of the stochastic volatility model with leverage to the S\&P 500 data.}
{\footnotesize
%{\tiny
\begin{tabular}{l | rr | rr | rr | rr | rr }\hline\hline
& \multicolumn{2}{c|}{Ac. Rate}  & \multicolumn{2}{c|}{Min. Inef.}
&\multicolumn{2}{c|}{Median Inef.}&\multicolumn{2}{c|}{Max. Inef.}
&\multicolumn{2}{c}{Median ECT} \\\cline{2-11}
\up{Algorithm}  & Median & IR & Median & IR & Median & IR & Median & IR & Median & IR \\\hline
&  \multicolumn{10}{c}{Standard Particle Filter}\\\hline
RWM3C   & 27.70 & 1.72 & 16.40 & 3.69 & 22.54 & 4.89 & 28.26 &  8.78 & 19.35 & 3.82 \\
IMH-MN  & 60.32 & 2.68 &  2.28 & 0.30 &  2.65 & 0.53 &  3.45 &  2.52 &  2.29 & 0.48 \\
\hline  & \multicolumn{10}{c}{Auxiliary Particle Filter}\\\hline
RWM3C   & 27.76 & 1.76 & 18.14 & 3.72 & 24.28 & 6.33 & 30.06 &  9.76 & 31.19 & 8.43 \\
IMH-MN  & 59.48 & 3.78 &  2.29 & 0.38 &  2.71 & 0.75 &  3.59 &  1.23 &  3.44 & 0.96 \\
\hline\hline
\end{tabular}
}
\label{table:SP500:data:MC:study}
%\end{sidewaystable}
\end{table}

We use importance sampling and bridge sampling to compute the marginal likelihoods of  the four SV models: the model with no leverage effect ($\rho = 0$) and no outlier effect ($\omega = 0$), the model that allows for leverage but not outliers, the model that allows for outliers but no leverage and the general model that allows for both outliers and leverage.
Table \ref{table:SV:model:comparison} shows the logarithms of the marginal likelihoods of the four models for a single run of each algorithm. The differences between the two approaches are very small. In this example, and based on our prior distributions, the SV model with leverage effects has the highest marginal likelihood.

\begin{table}[!ht]
\centering
\caption{Logarithms of the marginal likelihoods for four different SV models for the two \pf{}
algorithms computed using the \aimh{} algorithm. $BS$ and $IS$ mean bridge sampling and importance sampling.}
{\footnotesize
\begin{tabular}{ l | cc | cc}\hline\hline
          & \multicolumn{2}{c|}{Standard Particle Filter}
       &    \multicolumn{2}{c}{Auxiliary Particle Filter}\\\cline{2-5}
 \up{Model}   &	$\log(p_{BS}(y))$ & $\log(p_{IS}(y))$ &	$\log(p_{BS}(y))$ & $\log(p_{IS}(y))$ \\\hline
SV           & -1072.9    &   -1072.9 & -1072.9   &   -1072.9  \\
SV Lev.      & -1065.0    &   -1065.0 & -1065.0   &   -1065.0  \\
SV Out.      & -1076.6    &   -1076.6 & -1076.5   &   -1076.4  \\
SV Lev. Out. & -1069.3    &   -1069.3 & -1069.2   &   -1069.3  \\
\hline\hline
\end{tabular}
}
\label{table:SV:model:comparison}
\end{table}

We also ran a simulation using the second parallel computing method described in section \ref{ss: AS and parallel computation},
 using 10~000 iteration of the \aimh{}  samplers running on eight processors. Further implementation details are given in apppendix~\ref{app: S&V implement details}.
Table \ref{table:SP500:data:MC:study:parallel} summarizes the results. The table shows that the ECT of the \arwm{} algorithm is over 30 times larger than the ECT of the \aimh{} algorithm because the latter takes advantage of the parallelization.

%\begin{sidewaystable}[!ht]
\begin{table}[!ht]
\centering
\caption{Medians and interquartile range (IR) of the acceptance rates and the inefficiencies (minimum, median and maximum) and ECT = IF $\times$ time for ten iterations over twelve replications of the stochastic volatility model with leverage to the S\&P 500 data using the standard particle filter and parallel computing on eight processors.}
{\footnotesize
%{\tiny
\begin{tabular}{l | rr | rr | rr | rr | rr  }\hline \hline
& \multicolumn{2}{c|}{Ac. Rate}  & \multicolumn{2}{c|}{Min. Inef.}
&\multicolumn{2}{c|}{Median Inef.}&\multicolumn{2}{c|}{Max. Inef.}
&\multicolumn{2}{c}{Median ECT}  \\
%\cline{2-13}
\up{Algorithm}  & Median & IR & Median & IR & Median & IR & Median & IR & Median & IR  \\\hline
RWM3C   & 27.22 & 2.42 & 18.76 & 2.64 & 22.00 & 8.15 & 31.50 & 16.03 & 68.63 & 25.60  \\
IMH-MN  & 58.15 & 1.90 &  2.56 & 0.40 &  2.98 & 0.54 &  4.38 &  2.55 &  1.39 &  0.27 \\
\hline \hline
\end{tabular}
}
\label{table:SP500:data:MC:study:parallel}
\end{table}
%\end{sidewaystable}
%--------------------------------------------------------------------------------------------------
%--------------------------------------------------------------------------------------------------
% Negative Binomial model: Pitt and Walker data set. State-space AR(1) model
%--------------------------------------------------------------------------------------------------
%--------------------------------------------------------------------------------------------------
\subsection{Example 2: Negative binomial model}
\cite{pitt:walker:2005} consider the  following Poisson gamma model and give an MCMC method to estimate it.
\begin{align} \label{PW:pois:gamma:model}
\begin{split}
y_t \mid \omega_t     &\sim \mathcal{P}(\omega_t) \ , \quad
\omega_t \mid z_{t-1} \sim  \mathcal{G}(\nu+z_{t-1},\alpha+\beta)\ , \\
z_t\mid \omega_t      &\sim \mathcal{P}(\alpha\omega_t)\ , \quad
\omega_t              \sim  \mathcal{G}(\nu,\beta),
\end{split}
\end{align}
where $\alpha>0$, $\beta>0$ and $\nu>0$. We can integrate $\omega_t$ out in \eqref{PW:pois:gamma:model} to obtain the negative-binomial model
\begin{align}\label{PW:Neg:Bin:model}
\begin{split}
y_t \mid z_{t}   &\sim  \mathcal{NB}\left(\nu+z_{t},\dfrac{\alpha+\beta}{\alpha+\beta+1}\right)\ , \\
z_t \mid z_{t-1} & \sim  \mathcal{NB}\left(\nu+z_{t-1},\dfrac{\alpha+\beta}{2\alpha+\beta}\right)\ , \quad \text{with} \quad
z_t              \sim  \mathcal{NB}\left(\nu,\dfrac{\beta}{\alpha+\beta}\right).
\end{split}
\end{align}
We use the notation $\mathcal{P}(a)$ for a Poisson distribution with mean $a$, $\mathcal{G}(a,b)$ for a gamma distribution with shape parameter $a$, scale parameter $b$ and mean $a/b$ and $\mathcal{NB}(r,p)$ is a negative binomial distribution with $r$ number of successes,  $p$ the probability of success, and with mean
$r(1-p)/p$. One of the advantages of the approach of \cite{pitt:walker:2005}  is that it is easy to obtain the marginal distribution of $y_t$, which in this example is  $y_t\sim \mathcal{NB}(\nu,\beta/(\beta+1))$.
%--------------------------------------------------------------------------------------------------
%--------------------------------------------------------------------------------------------------
% Negative Binomial model: Weekly firearm homicides in Cape Town
%--------------------------------------------------------------------------------------------------
%--------------------------------------------------------------------------------------------------
\subsubsection{Weekly firearm homicides in Cape Town}\label{sss: weekly firearm homicides}
This section fits the negative binomial model~\eqref{PW:Neg:Bin:model} to the number of weekly firearm homicides {\citep[pp. 194-195]{mcdonald:zucchini:1997} in Cape Town from January 1, 1986 to December 31, 1991, ($T=313$ observations) shown in figure~\ref{plot:homicides}.
\begin{figure}[!t]
\centerline{\includegraphics[scale=0.4,angle=-90]{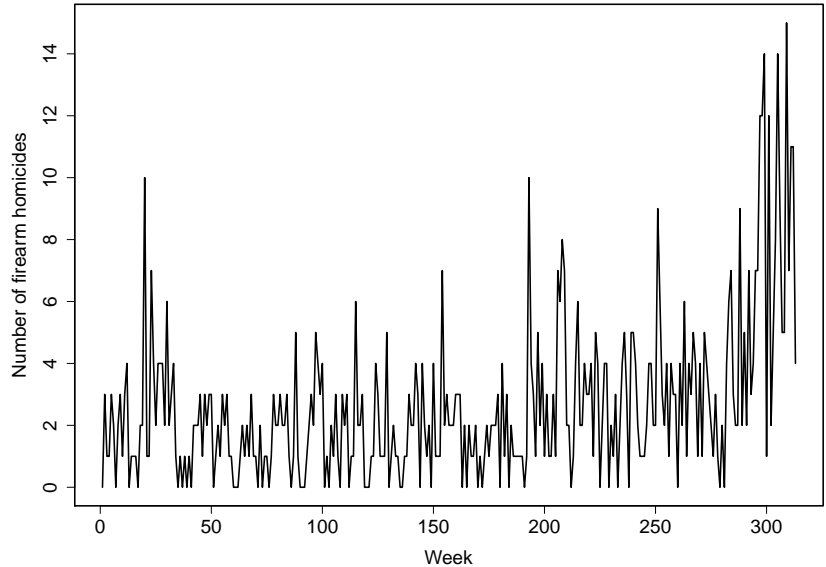}}
\caption{Number of weekly firearm homicides in Cape Town from January 1, 1986 to December 31, 1991.}
\label{plot:homicides}
\end{figure}
\cite{pitt:walker:2005} also fit this model to the data.
 We also fitted to this data a state-space Poisson model with a random walk transition equation,
\begin{align} \label{PW:Pois:RW:model}
y_t \mid \mu_t \sim& \mathcal{P}(\exp(\mu_t)) \ , \quad
\mu_{t}         =    \mu_{t-1} + \sigma\varepsilon_t,\hspace{0.5cm}\varepsilon_t\sim\mathcal{N}(0,1).
\end{align}
because figure~\ref{plot:homicides} suggests a possible nonstationarity in the data towards the end of the series.

The prior distributions for both model are based on our empirical analysis of the data. We assume that the parameters are independent a priori with prior distributions $\nu\sim\mathcal{HN}(25)$, $\beta\sim\mathcal{HN}(25)$ and $\alpha\sim\mathcal{HN}(400)$ for the negative binomial model, and $\sigma^2\sim\mathcal{HN}(1)$ and $\mu_0 \sim\mathcal{N}(0.4324,9)$ for the Poisson model, where $\mathcal{HN}(b^2)$ stands for a half-normal distribution with scale $b$.

Table~\ref{table:homicides:data:MC:study} presents the results of a Monte Carlo study using twelve replications with different random number seeds  for the negative binomial model. This simulation is based on the the first parallel computation method described in \ref{ss: AS and parallel computation}. The implementation details are given in appendix~\ref{app: weekly homicides}. The table shows that the inefficiencies of the \arwm{} are at least seven times as large as those of the \aimh{}.

%\begin{sidewaystable}[!ht]
\begin{table}[!ht]
\centering
\caption{Medians and interquartile range (IR) of the acceptance rates and the inefficiencies (minimum, median and maximum) and ECT = IF $\times$ time for twelve replications of the negative binomial model applied to the homicide data.}
{\footnotesize
%{\tiny
\begin{tabular}{l | rr | rr | rr | rr | rr }\hline\hline
& \multicolumn{2}{c|}{Ac. Rate}  & \multicolumn{2}{c|}{Min. Inef.}
&\multicolumn{2}{c|}{Median Inef.}&\multicolumn{2}{c|}{Max. Inef.}
&\multicolumn{2}{c}{Median ECT} \\\cline{2-11}
\up{Algorithm}  & Median & IR & Median & IR & Median & IR & Median & IR & Median & IR \\\hline
&  \multicolumn{10}{c}{Standard Particle Filter}\\\hline
RWM3C   & 25.32 & 2.26 & 16.79 & 3.47 & 21.79 & 4.85 & 27.33 & 12.39 & 31.95 &  8.14 \\
IMH-MN  & 62.90 & 2.68 &  2.24 & 0.77 &  2.65 & 0.83 &  3.82 &  3.35 &  4.18 &  1.33 \\
\hline  & \multicolumn{10}{c}{Auxiliary Particle Filter}\\\hline
RWM3C   & 24.58 & 1.78 & 19.03 & 5.81 & 23.58 & 5.61 & 33.81 & 10.07 & 37.00 &  8.76 \\
IMH-MN  & 56.72 & 4.74 &  2.42 & 1.07 &  2.93 & 1.27 &  3.82 &  3.38 &  5.08 &  2.40 \\
\hline\hline
\end{tabular}
}
\label{table:homicides:data:MC:study}
%\end{sidewaystable}
\end{table}

Table \ref{table:homicides:model:comparison} shows that the marginal likelihood of the negative binomial model is greater than that of the Poisson random walk model for the prior distributions chosen. A summary of the posterior distributions of the model parameters for the negative binomial model (not shown) provides similar results to those presented in \citet{pitt:walker:2005}.
\begin{table}[!ht]
\centering
\caption{Logarithms of the marginal likelihood estimates for the negative binomial and the Poisson models for the two \pf{}
 algorithms computed using the \aimh{} algorithm. $BS$ and $IS$ mean bridge sampling and importance sampling.}
{\footnotesize
\begin{tabular}{  l | cc | cc}\hline\hline
 & \multicolumn{2}{c|}{Standard Particle Filter}
         & \multicolumn{2}{c}{Auxiliary Particle Filter}\\\cline{2-5}
\up{Model}   &	$\log(p_{BS}(y))$ & $\log(p_{IS}(y))$ &	$\log(p_{BS}(y))$ & $\log(p_{IS}(y))$ \\\hline
 N. Binomial   & -620.6781 & -620.6582 & -620.5979 & -620.6767	\\
 Poisson       & -625.3963 & -625.3939 & -625.4304 & -625.4298	\\
\hline\hline
\end{tabular}
}
\label{table:homicides:model:comparison}
\end{table}

We also ran a simulation using the second parallel computing method described in
section~\ref{ss: AS and parallel computation},
 using 10~000 iteration of the \aimh{}  samplers running on eight processors. Implementation details are the same as for the second simulation in section~\ref{sss: sp}.
Table \ref{table:homicides:data:MC:study:parallel} summarizes the results and shows that the ECT of the \arwm{} algorithm is nearly 50 times larger than the ECT of the \aimh{} algorithm.
%\begin{sidewaystable}[!ht]
\begin{table}[!ht]
\centering
\caption{Medians and interquartile range (IR) of the acceptance rates and the inefficiencies (minimum, median and maximum) and ECT = IF $\times$ time for twelve replications of the negative binomial model applied to the homicides data using the standard particle filter and parallel computing on eight processors.}
{\footnotesize
%{\tiny
\begin{tabular}{l | rr | rr | rr | rr | rr  }\hline \hline
& \multicolumn{2}{c|}{Ac. Rate}  & \multicolumn{2}{c|}{Min. Inef.}
&\multicolumn{2}{c|}{Median Inef.}&\multicolumn{2}{c|}{Max. Inef.}
&\multicolumn{2}{c }{Median ECT}  \\
%\cline{2-13}
\up{Algorithm}  & Median & IR & Median & IR & Median & IR & Median & IR & Median & IR  \\\hline
RWM3C   & 24.68 &  1.96 & 18.46 & 6.40 & 25.72 & 8.13 & 36.14 & 11.89 & 103.14 & 33.89 \\
IMH-MN  & 52.52 & 12.64 &  2.98 & 1.05 &  3.37 & 1.69 &  4.41 &  2.64 &   2.04 &  1.01  \\
\hline \hline
\end{tabular}
}
\label{table:homicides:data:MC:study:parallel}
\end{table}
%--------------------------------------------------------------------------------------------------
%--------------------------------------------------------------------------------------------------
% State-space Poisson model with Fourier series seasonality representation
%--------------------------------------------------------------------------------------------------
%--------------------------------------------------------------------------------------------------
\subsection{Example 3: Poisson model} \label{ss: poisson model}
This section considers a state space model with a Poisson observation equation, dynamic level and slope equations as well as  explanatory variables
\begin{align} \label{eq:pois:model:seasonal}
\begin{split}
y_t      &\sim  \mathcal{P}(\exp(x_t \beta+\mu_t+s_t))\\
\mu_t    &=  \mu_{t-1} + a_{t-1} + \delta  I(t=t_{int}) + \sigma\varepsilon_{t},\quad          \varepsilon_{t}\sim\mathcal{N}(0,1),\\
 a_t      &=  a_{t-1} + \tau\xi_{t},\qquad\qquad\qquad\qquad\qquad\xi_{t}\sim\mathcal{N}(0,1),\\
s_t      &=  \sum_{j=1}^{J}\left\{\alpha_j\cos\left(\omega_j t\right)+\gamma_j\sin\left(\omega_j t\right)\right\},
\end{split}
\end{align}
where $\omega_t=2\pi j/h$ so that $s_t$ has period $h$. The variable $I(t=t_{int}) = 1 $ if $t = t_{int}$ and 0 otherwise so the model allows for a change in level in the $\mu_t$ equation if $\delta \neq 0 $.
We assume that the parameters are independent a priori with the following prior distributions:
$\beta  \sim \text{N}(0,\varphi_\beta^2\bfI)$,
$\mu_0    \sim \text{N}(\overline{\mu}_0,\varphi_\mu^2)$,
$a_0      \sim \text{N}(0,\varphi_a^2)$,
$\sigma^2 \sim \text{HN}(0,\varphi_{\sigma^2}^2)$,
$\tau^2   \sim \text{HN}(0,\varphi_{\tau^2}^2)$,
$\delta   \sim \text{N}(0,1)$,
$\alpha_j \sim \text{N}(0,\varphi_\alpha^2)$, and
$\gamma_j \sim \text{N}(0,\varphi_\gamma^2)$, for $j=1,\ldots,J$.
%--------------------------------------------------------------------------------------------------
%--------------------------------------------------------------------------------------------------
% Killed or injured children in Linz
%--------------------------------------------------------------------------------------------------
%--------------------------------------------------------------------------------------------------
\subsubsection{Killed or seriously injured children in Linz}\label{sss; ksi linz}
The first application of the Poisson model  is the number of children aged 6-10 that
were killed or seriously injured  by motor vehicles in Linz, Austria, from 1987 to 2002, corresponding to $T=192$ observations.
The data is analyzed  by \cite{fruhwirhschnatter:wagner:2006}.
 We fit the Poisson model
 at~\eqref{eq:pois:model:seasonal} to the data.  The seasonal pattern in our model uses a Fourier series representation
that differs from the state space model in~\cite{fruhwirhschnatter:wagner:2006}.
We also include the same explanatory variable $x_t=\log(z_t)$, where $z_t$ is the number of children living in Linz,
as used by \cite{fruhwirhschnatter:wagner:2006}. The coefficient $\beta$ at~\eqref{eq:pois:model:seasonal} is set to 1  so there  a multiplicative effect on the mean of the Poisson model. The hyperparameters in the prior are based on an empirical analysis of the data using the Ascombe {\citep{anscombe:1948}} transform. In particular, we set $\overline{\mu}_0=-8.3779$ , $\varphi_\mu^2 = 1.5$, $\varphi_a^2 = \varphi_\alpha^2 = \varphi_\gamma^2 = 0.005$, $\varphi_{\sigma^2}^2 = 0.2$, $\varphi_{\tau^2}^2 = 0.002$ and $\omega_j = 2\pi/12$.
An intervention parameter $I(t=t_{int})$ is included in the model to capture a possible decrease in the level of the series due to a change in the law in Linz in October 1, 1994 $(t_{int}=95)$ as can be seen in figure~\ref{plot:children}.
\begin{figure}[!ht]
\centerline{\includegraphics[scale=0.4,angle=-90]{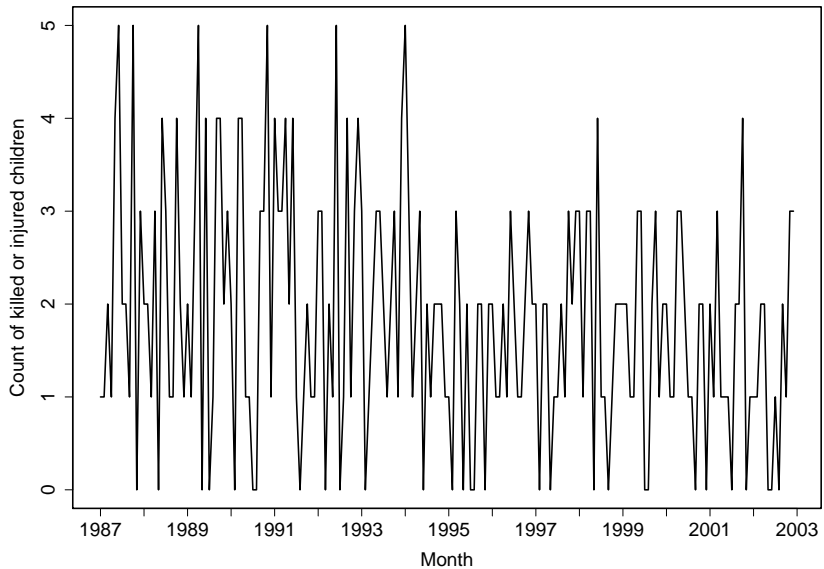}}
\caption{Monthly counts of killed or injured children from 1987 to 2002 in Linz.}
\label{plot:children}
\end{figure}

Table \ref{table:Linz:data:MC:study} presents the results of a Monte Carlo study using twelve replications with different random number seeds  for the Poisson model. This simulation is based on the the first parallel computation method described in \ref{ss: AS and parallel computation}. The implementation details are given in appendix~\ref{app: Linz ksi}. The table shows that the inefficiencies of the \arwm{} are at least seven times as large as those of the \aimh{}.

%\begin{sidewaystable}[!ht]
\begin{table}[!ht]
\centering
\caption{Medians and interquartile range (IR) of the acceptance rates and the inefficiencies (minimum, median and maximum) and ECT = IF $\times$ time for twelve replications of the level and trend state-space poisson model applied to the Linz data.}
{\footnotesize
%{\tiny
\begin{tabular}{l | rr | rr | rr | rr | rr }\hline\hline
& \multicolumn{2}{c|}{Ac. Rate}  & \multicolumn{2}{c|}{Min. Inef.}
&\multicolumn{2}{c|}{Median Inef.}&\multicolumn{2}{c|}{Max. Inef.}
&\multicolumn{2}{c}{Median ECT} \\\cline{2-11}
\up{Algorithm}  & Median & IR & Median & IR & Median & IR & Median & IR & Median & IR \\\hline
&  \multicolumn{10}{c}{Standard Particle Filter}\\\hline
RWM3C   & 25.71 & 1.88 & 40.61 & 5.22 & 56.41 & 6.11 & 81.04 &  9.57 & 14.54 & 1.44 \\
IMH-MN  & 40.90 & 5.23 &  4.15 & 0.88 &  6.04 & 2.56 & 15.90 & 29.70 &  1.55 & 0.66 \\
\hline  & \multicolumn{10}{c}{Auxiliary Particle Filter}\\\hline
RWM3C   & 27.12 & 1.77 & 40.38 & 7.52 & 55.94 & 5.27 & 80.31 & 22.59 & 21.76 & 2.09 \\
IMH-MN  & 41.78 & 3.80 &  4.16 & 0.72 &  6.47 & 2.14 & 22.03 & 28.13 &  2.43 & 0.74 \\
\hline\hline
\end{tabular}
}
\label{table:Linz:data:MC:study}
%\end{sidewaystable}
\end{table}

We compared the eight models given in table~\ref{table:Linz:model:selection} using marginal likelihood. All the models with seasonal effects include five harmonics. The table shows
that the simplest model is slightly better than the level and intervention model which is consistent with the results reported in \cite{fruhwirhschnatter:wagner:2006}. However, the intervention parameter is clearly negative with high probability and two of the seasonal coefficients have high probability of being different from zero (results not shown). The bridge and importance samplers give similar estimates of the marginal likelihoods.
\begin{table}[!ht]
\centering
\caption{Logarithms of the marginal likelihoods for different Poisson models for the two \pf{} algorithms.
$BS$ and $IS$ mean bridge sampling and importance sampling.}
{\footnotesize
\begin{tabular}{ l | cc | cc}\hline\hline
          & \multicolumn{2}{c|}{Standard Particle Filter}
         & \multicolumn{2}{c}{Auxiliary Particle Filter}\\\cline{2-5}
\up{Poisson model}   &	$\log(p_{BS}(y))$ & $\log(p_{IS}(y))$ &	$\log(p_{BS}(y))$ & $\log(p_{IS}(y))$ \\\hline
 Level                                      &-320.984 & -320.987 &-320.995 & -320.995 \\
 Level and trend                            &-333.110 & -333.122 &-333.100 & -333.100 \\
 Level and intervention                     &-321.279 & -321.278 &-321.285 & -321.291 \\
 Level, trend and intervention              &-333.867 & -333.866 &-333.861 & -333.857 \\
 Level and seasonality                      &-328.438 & -328.436 &-328.451 & -328.444 \\
Level, trend and seasonality               &-343.214 & -343.202 &-343.246 & -343.247 \\
 Level, intervention and seasonality        &-328.632 & -328.640 &-328.662 & -328.657 \\
Level, trend, intervention and seasonality &-341.253 & -341.246 &-341.267 & -341.264 \\
\hline\hline
\end{tabular}
}
\label{table:Linz:model:selection}
\end{table}

We also ran a simulation using the second parallel computing method described in section \ref{ss: AS and parallel computation}
with 20~000 iteration of the \aimh{}  sampler running on eight processors. Implementation details are in
appendix~\ref{app: Linz ksi}. Table \ref{table:Linz:data:MC:study:parallel}  summarizes the results and shows that the ECT of the \arwm{} algorithm is nearly 50 times larger than the ECT of the \aimh{} algorithm.
%\begin{sidewaystable}[!ht]
\begin{table}[!ht]
\centering
\caption{Medians and interquartile range (IR) of the acceptance rates and the inefficiencies (minimum, median and maximum) and ECT = IF $\times$ time for twelve replications of the level and trend state-space Poisson model applied to the Linz data using the standard particle filter and the parallel computing in eight processors.}
{\footnotesize
%{\tiny
\begin{tabular}{l | rr | rr | rr | rr | rr  }\hline \hline
& \multicolumn{2}{c|}{Ac. Rate}  & \multicolumn{2}{c|}{Min. Inef.}
&\multicolumn{2}{c|}{Median Inef.}&\multicolumn{2}{c|}{Max. Inef.}
&\multicolumn{2}{c|}{Median ECT}  \\
%\cline{2-13}
\up{Algorithm}  & Median & IR & Median & IR & Median & IR & Median & IR & Median & IR  \\\hline
RWM3C   & 26.37 & 0.86 & 40.60 & 7.57 & 56.41 & 4.40 & 85.81 & 26.95 & 107.70 & 13.41\\
IMH-MN  & 40.07 & 4.41 &  4.54 & 0.81 &  6.86 & 2.99 & 27.26 & 38.01 &   1.90 &  0.81 \\
\hline \hline
\end{tabular}
}
\label{table:Linz:data:MC:study:parallel}
\end{table}

%--------------------------------------------------------------------------------------------------
%--------------------------------------------------------------------------------------------------
% Sydney asthma data
%--------------------------------------------------------------------------------------------------
%--------------------------------------------------------------------------------------------------
\subsubsection{Sydney asthma data}\label{sss: asthma data}
 This example models the time series of daily counts of asthma presentations at the accident and emergency department of Campbelltown Hospital located in southwest metropolitan area of Sydney.
 figure~\ref{plot:asthma} is a plot of the data, which has 1461 observations from January 1, 1990 to December 31, 1994.
\cite{davis:dunsmuir:streett:2003} analyze this data using a Poisson model. \cite{davis:dunsmuir:streett:2003} argue that the peaks in the series can be lined up with the four terms in the school year
with the break between the first and second terms occurring at varying times because of the timing of the Easter vacation.
They include only one harmonic, $(\alpha\cos(2\pi t /365)+\gamma\sin(2\pi t /365))$ to model the seasonal effect, and model the peaks by constructing the explanatory variable
\begin{equation}\notag
P_{ij}(t) = p\left(\dfrac{t-T_{ij}}{100}\right),\hspace{0.25cm}\text{for}\hspace{0.25cm}i=1,2,3,4
\hspace{0.25cm}\text{and}\hspace{0.25cm}j=1,\ldots,1461
\end{equation}
where $T_{ij}$ is the start time for the $j$th school term in year $i$ and $p(x)\propto x^{a-1}(1-x)^{b-1}$
(a beta density), with parameter $a=2.5$ and $b=5$. There are sixteen such explanatory variables but their preliminary
analysis only includes eight of them corresponding to terms 1 and 2 across all four years. They also include the following explanatory variables: Sunday and Monday effects (dummy variables), maximum daily ozone, maximum daily NO2 and humidity.
\begin{figure}[!ht]
\centerline{\includegraphics[height=15cm,width=7.5cm,angle=-90]{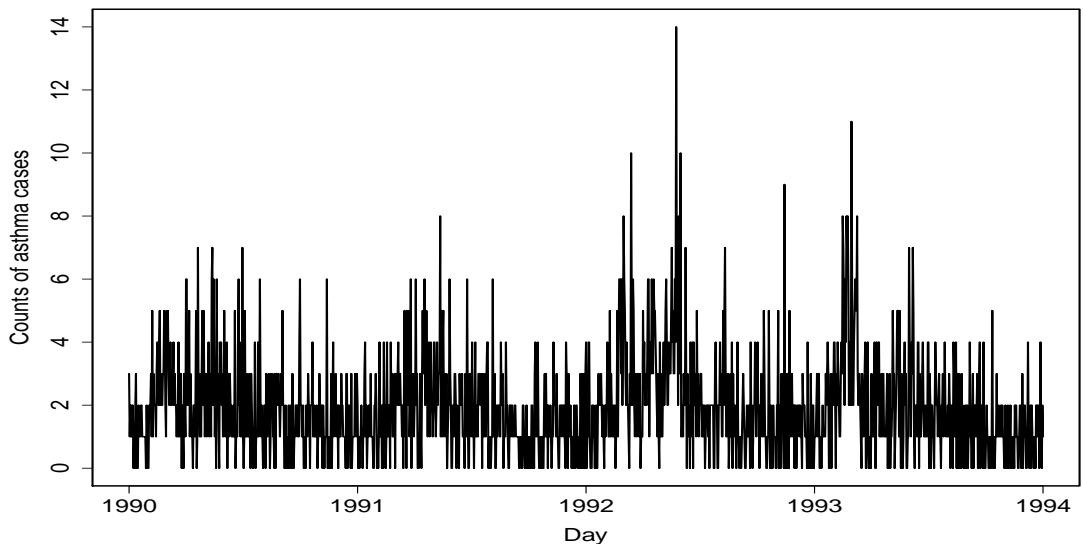}}
%\centerline{\includegraphics[scale=0.4,angle=-90]{asthmaplot.ps}}
\caption{Counts of asthma presentation in Campbelltown Hospital.}
\label{plot:asthma}
\end{figure}
We  apply model \eqref{eq:pois:model:seasonal} to the asthma data, but without the intervention variable, i.e. taking $\delta$ identically zero, and use the hyperparameters $\varphi_\beta^2=0.02$, $\overline{\mu}_0=0.5093$ , $\varphi_\mu^2=20$,
$\varphi_a^2=0.001$, $\varphi_{\sigma^2}^2=1.5$, $\varphi_{\tau^2}^2=0.002$,
$\varphi_\alpha^2=\varphi_\gamma^2=10$,  $\omega_j=2\pi/365$,
which are obtained through an empirical analysis of the data.
Table \ref{table:asthma:data:MC:study}  presents the results of a Monte Carlo study using twelve replications with different random number seeds  for the Poisson model with level, trend and seasonality, and the explanatory variables. This simulation is based on the first parallel computation method described in \ref{ss: AS and parallel computation},
with the implementation details given in appendix~\ref{app: asthma}.
The table shows that the inefficiencies of the \arwm{} are at least twice as large as those of the \aimh{}.
%\begin{sidewaystable}[!ht]
\begin{table}[!ht]
\centering
\caption{Asthma data: Medians and interquartile ranges (IR) of  acceptance rates and the inefficiencies (minimum, median and maximum) and ECT = IF $\times$ time for twelve replications of the level and slope state-space Poisson model. }
{\footnotesize
%{\tiny
\begin{tabular}{l | rr | rr | rr | rr | rr }\hline\hline
& \multicolumn{2}{c|}{Ac. Rate}  & \multicolumn{2}{c|}{Min. Inef.}
&\multicolumn{2}{c|}{Median Inef.}&\multicolumn{2}{c|}{Max. Inef.}
&\multicolumn{2}{c}{Median ECT} \\\cline{2-11}
\up{Algorithm}  & Median & IR & Median & IR & Median & IR & Median & IR & Median & IR \\\hline
&  \multicolumn{10}{c}{Standard Particle Filter}\\\hline
RWM3C   & 25.59 &  1.09 & 68.46 & 7.00 & 88.50 &  2.81 & 119.74 & 16.62 & 148.59 &  5.18 \\
IMH-MN  & 27.52 &  7.75 &  8.51 & 6.18 & 16.45 & 15.62 &  52.26 & 49.66 &  27.73 & 26.34 \\
\hline  & \multicolumn{10}{c}{General Auxiliary Particle Filter}\\\hline
RWM3C   & 26.52 &  3.45 & 73.96 &  8.06 & 87.41 &  1.64 & 114.99 & 14.29 & 219.95 &  4.26 \\
IMH-MN  & 18.45 &  7.22 & 14.52 &  7.32 & 26.16 & 11.38 &  55.21 & 16.44 &  66.03 & 28.71 \\
\hline\hline
\end{tabular}
}
\label{table:asthma:data:MC:study}
%\end{sidewaystable}
\end{table}

Table \ref{table:asthma:model:selection} uses marginal likelihood to compare the model
with just the level $\mu_t$ in the transition equation to a model containing the level $\mu_t$ and the trend $a_t$, with the
simpler model preferred.
\begin{table}[!ht]
\centering
\caption{Logarithms of the marginal likelihood estimates for the two Poisson models estimated using the two particle filters.
$BS$ and $IS$ mean bridge sampling and importance sampling.}
{\footnotesize
\begin{tabular}{  l | cc | cc}\hline\hline
 & \multicolumn{2}{c|}{Standard Particle Filter}
         & \multicolumn{2}{c}{Auxiliary Particle Filter}\\\cline{2-5}
\up{Model}   &	$\log(p_{BS}(y))$ & $\log(p_{IS}(y))$ &	$\log(p_{BS}(y))$ & $\log(p_{IS}(y))$ \\\hline
Level and seasonality        & -2558.3 & -2558.3 & -2558.2 & -2558.3 \\
 Level, trend and seasonality & -2577.3 & -2577.3 & -2577.2 & -2577.2 \\
\hline\hline
\end{tabular}
}
\label{table:asthma:model:selection}
\end{table}
\section*{Acknowledgment}
The research of Robert Kohn and Ralph S. Silva was partially supported by an ARC Discovery Grant DP0667069. We thank Professor Sylvia Fr\"uhwirth-Schnatter for the Linz data and Professor William T. M. Dunsmuir for the asthma data.
%--------------------------------------------------------------------------------------------------
%--------------------------------------------------------------------------------------------------
% Bibliography
%--------------------------------------------------------------------------------------------------
%--------------------------------------------------------------------------------------------------
\bibliographystyle{asa}
\bibliography{pf}

\begin{appendices}
\section{Standard particle filter}\label{A: regular PF}
This section outlines the
standard Sampling-Importance-Resampling (SIR) particle filter of \cite%
{gordon:salmond:smith:1993}. We suppress the dependence on the fixed
parameter $\theta $ for notational convenience. Suppose that
that we have samples\textit{\ }$x_{t-1}^{k}\sim p(x_{t-1}|y_{1:t-1})$ for $%
k=1,...,M$. The particle filter works by taking this sample, from the
filtering density at time $t-1$, and translating it into a sample from the
filtering density at time $t$. The first step involves simply passing each
of these samples through the transition density to obtain $%
\widetilde{x}_{t}^{k}\sim p(x_{t}|x_{t-1}^{k})$, for $k=1,...,M$, which
produces samples which are approximately distributed from equation %
\eqref{eq: update1}. These samples $\{\widetilde{x}_{t}^{k}\}$ are therefore
samples from the prediction density
(we denote filtered samples as $x$ and the corresponding predictive samples
as $\widetilde{x}$).  To each of these samples, for $k=1,...,M$, we attach
the following weights, $\omega _{t}^{k},$ and corresponding masses, $\pi
_{t}^{k},$
\begin{equation}
\omega _{t}^{k}=p(y_{t}|\widetilde{x}_{t}^{k}),\text{ \ \ \ \ \ \ }\pi
_{t}^{k}=\frac{\omega _{t}^{k}}{\sum_{i=1}^{M}\omega _{t}^{i}}.
\label{eq:PF:update:probab}
\end{equation}%
This collection $\left\{ (\widetilde{x}_{t}^{k},\pi _{t}^{k})\right\}
_{k=1}^{M}$ is now a discrete approximation to the filtering density $%
p(x_{t}|y_{1:t})$. Explicitly, we may write this approximation, in terms of
Dirac-delta functions, $\delta (\cdot )$, as,%
\begin{equation}
\widehat{p}(x_{t}|y_{1:t})=\sum_{k=1}^{M}\pi _{t}^{k}\delta (x_{t}-%
\widetilde{x}_{t}^{k}).  \label{eq:PF:mixture_rep}
\end{equation}%
We need to resample from this mass function to obtain an equally
weighted sample. However, prior to doing this we may estimate the term %
\eqref{eq: one term likelihood} unbiasedly \citep{delmoral:2004} by the
denominator at equation~\eqref{eq:PF:update:probab},%
\begin{equation}
\frac{1}{M}\sum_{k=1}^{M}p(y_{t}|\widetilde{x}_{t}^{k})=\frac{1}{M}%
\sum_{k=1}^{M}\omega _{t}^{k}.  \label{eq:RPF:likelihood}
\end{equation}%
We may also estimate any moments under the filtering density, say $%
E[g(x_{t})|y_{1:t}]$, in the Rao-Blackwellised form as,%
\begin{equation*}
\text{\ \ \ }\sum_{k=1}^{M}g(\widetilde{x}_{t}^{k})\pi _{t}^{k}\text{.}
\end{equation*}%
Typically these estimators are more efficient than using the resampled
analogs.

To produce an equally weighted sample from equation~\eqref{eq:PF:mixture_rep}, we need
only think about the sampling of the discrete univariate index $k$ with mass
$\pi _{t}^{k}$ for $k=1,...,M$. This is a multinomial sample and is the
equivalent of a weighted bootstrap.\footnote{%
The SIR filter is sometimes referred to as the bootstrap filter for this
reason.} We then have a sample $z_{t}^{1},...,z_{t}^{M}$ of resampled
indices. Having sampled in this manner, we can now associate our resampled
points, which we call $x_{t}^{k}$ for $k=1,...,M$, with the predictive
points,%
\begin{equation*}
x_{t}^{k}=\widetilde{x}_{t}^{z_{t}^{k}}\text{ for }k=1,...,M.
\end{equation*}%
The method now proceeds to the next time step in a similar fashion.

We may replace the multinomial resampling (weighted bootstrap) procedure
with a stratified sampling step instead. This does not affect the validity
of the particle filter or the subsequent MCMC\ strategy that we pursue.

\section{Proof of theorem \ref{thm: convergence apf}}\label{A: APF}
This section briefly outlines the conditions for theorem~\ref{thm: convergence apf} to hold and its proof.
The generic \apf{} of \citet{pitt:shephard:1999} uses an importance density of the form
$p(y_t|z_t^k;\theta)p(x_{t}|x_{t-1}^k)$, where
$z_t^k$ is a suitable value of $x_t$. We replace the term  $p(y_t|z_t^k;\theta)$ in the importance density
by $\veps \phi_t + (1-\veps) p(y_t|z_t^k;\theta)$ where $\phi_t$ is defined in theorem~\ref{thm: convergence}.
By \cite{pitt:2002}, the term \eqref{eq: one term likelihood} is estimated unbiasedly by
\begin{align*}
\phat(y_t|y_{1:t-1}; \theta) & = \biggl ( \frac1M \sum_{k=1}^M p(y_t|x_t^k; \theta ) \biggr ) \biggl ( \frac1M \sum_{k=1}^M \frac { p(y_t|x_t^k; \theta )}
{\veps \phi_t +(1-\veps) p(y_t|z_t^k;\theta)}\biggr )\ .
\end{align*}
It follows that $\phat(y_t|y_{1:t-1}; \theta) \leq \phi_t$. The rest of the proof is the same as that of Theorem~\ref{thm: convergence}.

\section{Adaptive sampling schemes} \label{app: adaptive sampling}
This appendix describes the two adaptive sampling schemes used in the paper.
\subsection{Adaptive random walk Metropolis}\label{subsection:ARWM}
The adaptive random walk Metropolis proposal of \cite{roberts:rosenthal:2008} is
\begin{equation} \label{e:arwm proposal}
q_j(\theta;\theta_{j-1}) =\omega_{1j}\phi_d (\theta; \theta_{j-1}, \kappa_1\Sigma_1) + \omega_{2j}\phi_d (\theta; \theta_{j-1}, \kappa_2\Sigma_{2j})
\end{equation}
where $d$ is the dimension of $\theta$ and $\phi_d(\theta; \thetat, \Sigma)$ is a multivariate
$d$ dimensional normal density in $\theta$ with mean $\thetat$ and covariance matrix $\Sigma$.
In \eqref{e:arwm proposal}, $\omega_{1j} = 1$ for $j \leq j_0$, with $j_0$ representing the initial iterations,
$\omega_{1j} = 0.05$ for $j > j_0$ with $\omega_{2j} = 1 -  \omega_{1j}$; $\kappa_1 = 0.1^2/d, \kappa_2 = 2.38^2/d, \Sigma_1 $ is a constant covariance matrix, which is taken as the identity matrix by \cite{roberts:rosenthal:2008} but can be based on the Laplace approximation or some other estimate. The matrix $\Sigma_{2j}$ is the sample covariance matrix of the first $j-1$ iterates. The scalar $\kappa_1$ is meant to achieve a high acceptance rate by moving the sampler locally, while the scalar $\kappa_2$ is considered to be optimal \citep{roberts:gelman:gilks:1997} for a random walk proposal when the target is a multivariate normal. We note that the acceptance probability \eqref{e:adaptive accep prob} for the adaptive random walk Metropolis simplifies to
\begin{equation} \label{e:arwm accep prob}
\alpha(\theta_{j-1},u_{j-1};\theta_{j}^p,u^{p}) = \min \biggl \{1,\frac{p(y|\theta_{j}^p,u_j^p)p(\theta^p)}
{p(y|\theta_{j-1},u_{j-1})p(\theta_{j-1})}\biggr \} \ .
\end{equation}
We refine the two component random walk Metropolis proposal in \eqref{e:arwm proposal} by adding a third component with
$\Sigma_{3j}=\Sigma_{2j}$ and with $\kappa_3 = 25 \gg \kappa_1, \kappa_2$. We take $\omega_{3j} = 0$ if $j \leq j_0 $, $\omega_{3j} = 0.05 $ for $j > j_0$ and $\omega_{2j} = 1- \omega_{1j} - \omega_{3j}$. We refer to this proposal as the three component adaptive random walk. The purpose of the heavier tailed third component is to allow the sampler to explore the state space more effectively by making it easier to leave local modes.

%--------------------------------------------------------------------------------------------------
%--------------------------------------------------------------------------------------------------
% A mixture of normals based proposal density
%--------------------------------------------------------------------------------------------------
%--------------------------------------------------------------------------------------------------
\subsection{ A proposal density based on a mixture of normals}\label{subsection:AIMH:MN}
The proposal density of the adaptive independent Metropolis-Hastings approach of \cite{giordani:kohn:2008} is a mixture with four terms of the form
\begin{equation}\label{eq:aimh proposal}
q_j(\theta)=\sum_{k=1}^4\omega_{kj}g_{k}(\theta| \lambda_{kj} )\,
\quad  \quad \omega_{kj} \ge 0, \quad\text{for} \quad k=1, \dots, 4
\quad \text{and}\quad  \sum_{k=1}^4 \omega_{kj} = 1\ ,
\end{equation}
with $\lambda_{kj}$ the  parameter vector for the density $g_{kj}(\theta;\lambda_{kj})$.
The sampling scheme is run in two stages, which are described below. Throughout each stage, the parameters in the
first two terms are kept fixed.  The first term $g_{1}(\theta| \lambda_{1j})$ is an estimate of the target density and
the second term $g_{2}(\theta| \lambda_{2j})$ is a heavy tailed version of $g_{1}(\theta| \lambda_{1j})$.
The third term $g_{3}(\theta| \lambda_{3j})$ is an estimate of the target that is updated or adapted as the simulation
progresses and the fourth term $g_{4}(\theta| \lambda_{4j})$ is a heavy tailed version of the third term.
In the first stage $g_{1j}(\theta; \lambda_{1j})$ is a Gaussian density constructed from a preliminary run,
of the three component adaptive random walk. Throughout, $g_{2}(\theta| \lambda_{2j})$ has
the same component means and probabilities as $g_{1}(\theta| \lambda_{1j})$, but its component covariance matrices are ten times those of $g_{1}(\theta| \lambda_{1j})$. The term $g_{3}(\theta| \lambda_{3j})$ is a mixture of normals
and $g_{4}(\theta| \lambda_{4j})$ is also a mixture of normals obtained by taking its component probabilities and means equal to those of $g_{3}(\theta| \lambda_{3j})$, and its component covariance matrices equal to 20 times those of $g_{3}(\theta| \lambda_{3j})$. The first stage begins by using $g_{1}(\theta|\lambda_{1j})$ and $g_{2}(\theta| \lambda_{2j})$  only with,  for example, $\omega_{1j} = 0.8$ and $\omega_{2j} = 0.2$, until there is a sufficiently large number of iterates to form $g_{3}(\theta| \lambda_{3j})$. After that we set $\omega_{1j} = 0.15, \omega_{2j} = 0.05, \omega_{3j} = 0.7 $ and $\omega_{4j} = 0.1$. We begin with a single normal density for $g_{3}(\theta| \lambda_{3j})$ and as the simulation progresses we add more components up to a maximum of six according to a schedule that depends on the ratio of the number of accepted draws to the dimension of $\theta$.

In the second stage, $g_{1}(\theta| \lambda_{1j})$ is set to the value of $g_{3}(\theta| \lambda_{3j})$ at the end of the first stage and $g_{2}(\theta| \lambda_{2j})$ and $g_{4}(\theta| \lambda_{4j})$ are constructed as described above. The heavy-tailed densities $g_{2}(\theta| \lambda_{2j})$ and $g_{4}(\theta| \lambda_{4j})$ are included as a
defensive strategy to get out of  local modes and to explore the sample space of the target distribution more effectively.

It is computationally too expensive to update $g_{3}(\theta| \lambda_{3j})$ (and hence $g_{4}(\theta| \lambda_{4j})$) at every iteration so we update them according to a schedule that depends on the problem and the size of the parameter vector.
\section{Marginal likelihood evaluation using bridge and importance sampling} \label{A: BS and IS}
Suppose that $q(\theta)$ is an approximation to $p(\theta| y )$ which can be evaluated explicitly. Bridge sampling \citep{meng:wong:1996} estimates the marginal likelihood as follows. Let
\begin{equation}\notag
t(\theta)=\left(\dfrac{p(y|\theta)p(\theta)}{U}+q(\theta)\right)^{-1},
\end{equation}
where $U$ is a positive constant. Let
\begin{align} \label{eq: bridge1}
\begin{split}
A & = \int t(\theta)q(\theta)p(\theta\mid y)d\theta \ . \qquad \text{Then,} \\
A &  = \frac{A_1}{p(y)} \qquad \text{where} \qquad A_1  = \int t(\theta)q(\theta)p(y\mid \theta) p(\theta) d\theta \ .
\end{split}
\end{align}
Suppose the sequence of iterates $\{\theta^{(j)},j=1,\ldots,M\}$ is generated from the posterior density $p(\theta|y)$ and a second sequence of iterates $\{\ttheta^{(k)},k=1,\ldots,M\}$ is generated from $q(\theta)$. Then
\begin{align}\notag
\widehat{A} & = \dfrac{1}{M}\sum_{j=1}^{M}t(\theta^{(j)})q(\theta^{(j)}), \quad
\widehat{A}_1  =\dfrac{1}{M}\sum_{k=1}^{K} t(\theta^{(k)})p(y|\theta^{(k)})p(\theta^{(k)})\quad \text{and}\quad
\widehat{p}_{BS}(y)=\dfrac{\widehat{A}_1}{\widehat{A}}
\end{align}
are estimates of $A$ and $A_1$ and $\widehat{p}_{BS}(y)$ is the bridge sampling estimator of the marginal likelihood $p(y)$.

In adaptive sampling, $q(\theta)$ is  the mixture of normals proposal.
Although $U$ can be any positive constant, it is more efficient if $U$ is a reasonable estimate of $p(y)$. One way to do so is to take
$
\Uhat  = p(y|\theta^*)p(\theta^*)/q(\theta^*)$,
where $\theta^*$ is the posterior mean of $\theta$ obtained from the posterior simulation.

An alternative method to estimate of the marginal likelihood $p(y)$ is to use importance sampling based on the proposal distribution $q(\theta)$ \citep{geweke:1989,chen:shao:1997}. That is,
\begin{equation}\nonumber
\widehat{p}_{IS}(y)=\dfrac{1}{K}\sum_{k=1}^{K}\dfrac{p(y|\theta^{(k)})p(\theta^{(k)})}{q(\theta^{(k)})}.
\end{equation}
Since our proposal distributions have at least one heavy tailed component, the importance sampling ratios are likely to be bounded and well-behaved, as in the examples in this paper.

\section{Implementation details and sampling schedules} \label{app: Implementation}
We coded most of the algorithms in MATLAB, with a small proportion of the code
written using C/Mex files. For the particle filters, we also use a C/Mex file for the resampling step using an efficient algorithm to draw from a general discrete distribution \citep{walker:1977} available as a C function in the GNU Scientific Library \citep{GSL:2009}. We carried out the estimation  on an SGI cluster with 42 compute nodes. Each of them is an SGI Altix XE320 with two Intel Xeon X5472
(quad core 3.0GHz) CPUs with at least 16GB memory. We ran parallel jobs using up to eight processors and MATLAB 2009.

\subsection{Implementation details for the S\&P 500 index data} \label{app: S&V implement details}
This section gives the implementation details for the first and second simulations in section~\ref{sss: sp}. The
first simulation uses a sampling rate of $M=3000$ particles for each time period for each of eight processor, so each step of the
particle filter uses 24~000 particles. The number of iterations of the adaptive samplers is 10 000 with the updates of the proposal distributions for the adaptive independent Metropolis-Hastings samplers performed at iterations 100, 200, 500, 1~000, 2~000, 3~000, 4~000, 5~000, 6~000 and 7~500. The \aimh{} sampling scheme is initialized using 2000 iterations of the ARWM3C. The initial proposal for all the AIMH algorithms are based on a multivariate normal distribution estimated from these draws and the initial starting values are the sample means.

The second simulation uses eight processors with block sizes for each processor of
15, 25, 60, 125, 250, 375, 500, 625, 750 and 940,  corresponding to
120, 200, 480, 1~000, 2~000, 3~000, 4~000, 5~000, 6~000 and 7~520 proposed parameter values
before each update of the proposal density.
The  proposal distribution is updated at the completion of each block.
In this simulation $M=10~000$ for the standard particle filter.

\subsection{Implementation details for the weekly homicide data} \label{app: weekly homicides}
This section gives the implementation details for the first and second simulations in
section~\ref{sss: weekly firearm homicides}.  The sampling rate for the first simulation is
$M= 2~500$ particles on each of eight parallel processors, so each step of the
particle filter uses 24~000 particles. The number of iterations of the adaptive sampling
algorithms is set to 10~000 with the updates of the proposal distributions for the adaptive independent Metropolis-Hastings samplers performed at iterations 100, 200, 500, 1~000, 2~000, 3~000, 4~000, 5~000, 6~000, and 7~500.
The simulation is initialized as in \ref{app: S&V implement details}.

The schedule for the second simulation is the same as in section~\ref{app: S&V implement details}.

\subsection{Implementation details for the analysis of the Linz data} \label{app: Linz ksi}
This section gives the implementation details for the first and second simulations in
section~\ref{sss; ksi linz}. The sampling  rate for the first simulation is
$M= 4~500$ particles in each of eight parallel processors.
The number of iterations is 20 000 with updates of the proposal distribution
for the adaptive independent Metropolis-Hastings sampler performed
at iterations 300, 1~000, 3~000, 5~000, 10~000, and 15~000.
The simulation is initialized as in \ref{app: S&V implement details}.

For the second simulation the number of iterations of the adaptive samplers is 20 000 with the adaptive independent Metropolis-Hastings sampler running on eight processors with the block sizes for each processor
40, 125, 375, 625, 1250, and 1875, corresponding to
320, 1000, 480, 3~000, 5~000, 10~000 and 15~000 proposed parameter values.
The updates of the proposal distribution occur at the end of each block.
In this simulation, $M= 30~000$ particles for the standard particle filter.

\subsection{Implementation details for analysis of the asthma data} \label{app: asthma}
This section gives the implementation details for the simulation in
section~\ref{sss: asthma data}. The sampling  rate for the simulation is
$M= 4~000$ particles in each of eight parallel processes. The number of iterations is set
to 50 000 with the updates of the proposal distributions for the adaptive independent Metropolis-Hastings
samplers performed at iterations
~1~000, 2~000, 3~000, 4~000, 5~000, 7~000, 8~000, 10~000, 15~000, 20~000, 25~000, 30~000, and 40~000.
\end{appendices}

\end{document}